\documentclass{PoS}

\usepackage{epsf}
\usepackage{latexsym,times,mathptmx,amssymb,euscript}
\usepackage{cite}
\usepackage[italian,english]{babel}
\usepackage[T1]{fontenc}
\usepackage{graphicx,xcolor}
\usepackage[format=hang,font=small]{caption}
\usepackage{amsmath}
\usepackage{amsfonts}
\usepackage{float}
\usepackage{amssymb}
\usepackage{booktabs}

\newcommand{\beq}[1]{
\begin{equation}\label{#1}}
\newcommand{\eeq}{\end{equation}}
\newcommand{\bea}[1]{
\begin{eqnarray}\label{#1}}
\newcommand{\eea}{\end{eqnarray}}
\newcommand{\barr}{
\begin{array}}
\newcommand{\earr}{\end{array}}
\newcommand{\drv}{{\rm d}}
\newcommand{\vsb}{\vspace{-0.10cm}}

\title{High-energy effects in forward inclusive dijet and hadron-jet production}

\ShortTitle{High-energy effects in forward inclusive dijet and hadron-jet production}

\author{Andr\`ee Dafne Bolognino\\
  Universit\`a della Calabria \& INFN - Gruppo collegato di Cosenza, \\
  I-87036 Arcavacata di Rende, Cosenza, Italy\\
  E-mail: \email{ad.bolognino@unical.it}}

\author{\speaker{Francesco Giovanni Celiberto}\\
  Universit\`a di Pavia \& INFN - Sezione di Pavia, I-27100 Pavia, Italy\\
  E-mail: \email{francescogiovanni.celiberto@unipv.it}}

\author{Dmitry Yu. Ivanov\\
  Sobolev Institute of Mathematics, 630090 Novosibirsk, Russia\\
  Novosibirsk State University, 630090 Novosibirsk, Russia\\
  E-mail: \email{d-ivanov@math.nsc.ru}}

\author{Mohammed M.A. Mohammed\\
  Universit\`a della Calabria \& INFN - Gruppo collegato di Cosenza, \\
  I-87036 Arcavacata di Rende, Cosenza, Italy\\
  E-mail: \email{mohammed.maher@unical.it}}

\author{Alessandro Papa\\
  Universit\`a della Calabria \& INFN - Gruppo collegato di Cosenza,\\
  I-87036 Arcavacata di Rende, Cosenza, Italy\\
  E-mail: \email{alessandro.papa@fis.unical.it}}

\abstract{
Pursuing the goal to single out the validity region of the high-energy resummation, better known as BFKL approach, and to possibly disentangle BFKL effects from the ones coming from a DGLAP-inspired, fixed-order description, new predictions for the forward inclusive hadron-jet production, tailored on the CMS and CASTOR acceptances, are given.
}

\FullConference{XXVII International Workshop on Deep-Inelastic Scattering and Related Subjects - DIS2019\\
		8-12 April, 2019\\
		Torino, Italy}

\begin{document}

\section{Introduction}
\vsb

The Balitsky-Fadin-Kuraev-Lipatov (BFKL)~\cite{BFKL} approach represents the most effective mechanism to resum to all orders, both in the leading (LLA) and the next-to-leading (NLA) approximation, large energy logarithmic contributions rising in the \emph{Regge limit} of QCD. 
With the aim of enhancing our knowledge of this high-energy regime, a wide range of semi-hard channels~\cite{Gribov:1984tu} (see~\cite{Celiberto:2017ius} for some applications) has been suggested so far: the exclusive electroproduction of one~\cite{Bolognino:2018rhb,Bolognino:2018mlw,Bolognino:2019bko} or two light vector mesons~\cite{Ivanov:2004pp,Ivanov:2005gn,Ivanov:2006gt,Enberg:2005eq}, the inclusive hadroproduction of two jets with large transverse momenta well separated in rapidity (Mueller--Navelet reaction~\cite{Mueller:1986ey}), for which a wealth of phenomenological analyses have been carried out so far~\cite{Colferai:2010wu,Caporale:2012ih,Ducloue:2013wmi,Ducloue:2013bva,Caporale:2013uva,Ducloue:2014koa,Caporale:2014gpa,Ducloue:2015jba,Caporale:2015uva,Celiberto:2015yba,Celiberto:2015mpa,Celiberto:2016ygs,Caporale:2018qnm,Chachamis:2015crx}, the inclusive emission of two charged light hadrons~\cite{Ivanov:2012iv,Celiberto:2016hae,Celiberto:2017ptm}, multi-jet production~\cite{Caporale:2015vya,Caporale:2015int,Caporale:2016soq,Caporale:2016xku,Celiberto:2016vhn,Caporale:2016pqe,Caporale:2016zkc}, heavy-quark pair photo- and hadroproduction~\cite{Celiberto:2017nyx}, $J/\Psi$-jet~\cite{Boussarie:2017oae} correlations and forward Drell--Yan dilepton production~\cite{Motyka:2014lya,Brzeminski:2016lwh,Celiberto:2018muu} with a possible backward-jet emission~\cite{Golec-Biernat:2018kem,Deak:2018obv}.

We propose in this work the study of a novel semi-hard reaction in the NLA BFKL approach, the inclusive hadron-jet production~\cite{Bolognino:2018oth,Bolognino:2019yqj}, {\it i.e.} when a charged light hadron and a jet are emitted in the final state with large transverse momenta, $\kappa_{H,J}$, and featuring a strong separation in rapidity. This analysis presents new, interesting features. On one hand, the concurrent detection of two basically distinct objects leads to a natural asymmetric configuration in the $\kappa$-plane, thus permitting to better discriminate pure BFKL effects from DGLAP ones. On the other hand, considering just one hadron in the final state, instead of two ones, allow us to dampen ``minimum-bias'' effects, thus easing to match experimental data. Finally, this process serves as a testfield for the comparison of different jet algorithms and parameterizations for fragmentation functions (FFs).

\section{Theoretical setup and numerical analysis}
\vsb

The reaction under exam is
\begin{eqnarray}
\label{process}
{\rm proton}(p_1) + {\rm proton}(p_2) 
\to 
{\rm hadron}(\kappa_H, y_H) + {\rm X} + {\rm jet}(\kappa_J, y_J) \;,
\end{eqnarray}
where a charged light hadron ($\pi^{\pm}, K^{\pm}, p \left(\bar p\right)$), and a
jet, with large transverse momenta, $\kappa_{H,J} \gg \Lambda_{\rm QCD}$, and featuring a large separation in rapidity, $Y \equiv y_H - y_J$, are detected together with a secondary, inclusive hadronic system, $X$. We can write the differential cross section as
\begin{equation}
\frac{\drv \sigma}
{\drv y_H \drv y_J\, \drv |\vec \kappa_H| \, \drv |\vec \kappa_J| \drv \vartheta_H \drv \vartheta_J}
=\frac{1}{(2\pi)^2}\left[{\cal C}_0+\sum_{n=1}^\infty  2\cos (n\vartheta )\,
{\cal C}_n\right]\, ,
\end{equation}
with $\vartheta \equiv \vartheta_H - \vartheta_J - \pi$, $\vartheta_{H,J}$ being the hadron/jet azimuthal angle. Here, ${\cal C}_0$ represents the $\vartheta$-averaged cross section, whereas the ${\cal C}_{n > 0}$ coefficients carry out information about the hadron-jet azimuthal distribution. With the aim of matching kinematic configurations typical of LHC analyses, we take the \emph{integrated} coefficients over the final-state phase space and keep fixed the rapidity interval, $Y$, between the hadron and the jet: 
\begin{equation}
\label{Cn_int}
C_n= 
\int_{y^{\rm min}_H}^{y^{\rm max}_H} \drv y_H
\int_{y^{\rm min}_J}^{y^{\rm max}_J} \drv y_J\int_{k^{\rm min}_H}^{k^{\rm max}_H} \drv \kappa_H
\int_{k^{\rm min}_J}^{{k^{\rm max}_J}} \drv \kappa_J
\, \delta \left( Y - ( y_H - y_J ) \right)
\, {\cal C}_n 
\, .
\end{equation}

\begin{figure}[t]
  \centering
  \includegraphics[scale=0.31,clip]{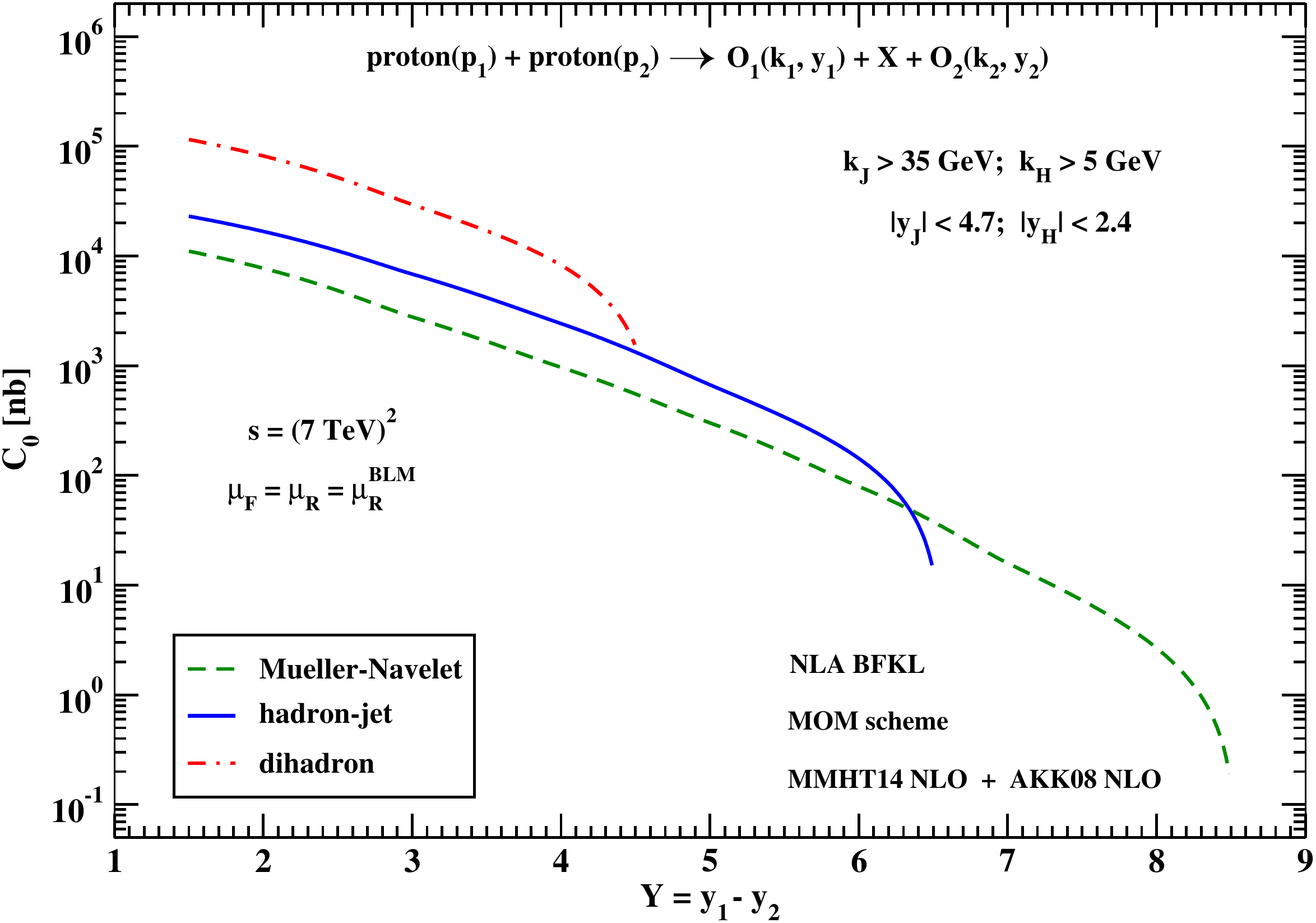}
  \hspace{0.15cm}
  \includegraphics[scale=0.31,clip]{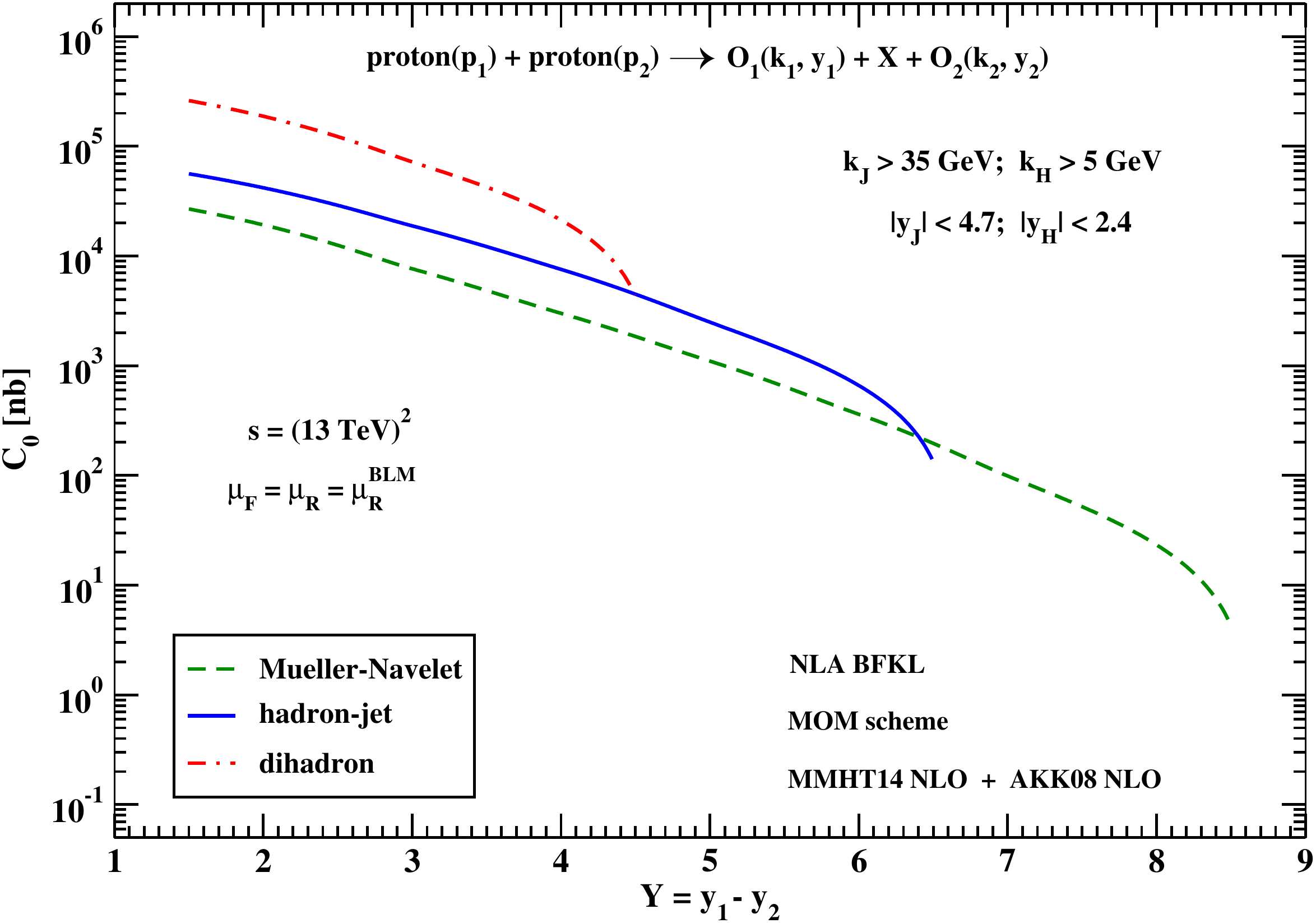}	
  \caption{$C_0$ in different inclusive NLA BFKL reactions, for $\sqrt{s} = 7$, 13 TeV in the {\it CMS-jet} configuration.}
  \label{fig:C0_comp_NLA_BLM_CMS}
\end{figure}

\begin{figure}[p]
 \centering

  \vspace{-0.35cm}

  \includegraphics[scale=0.30,clip]{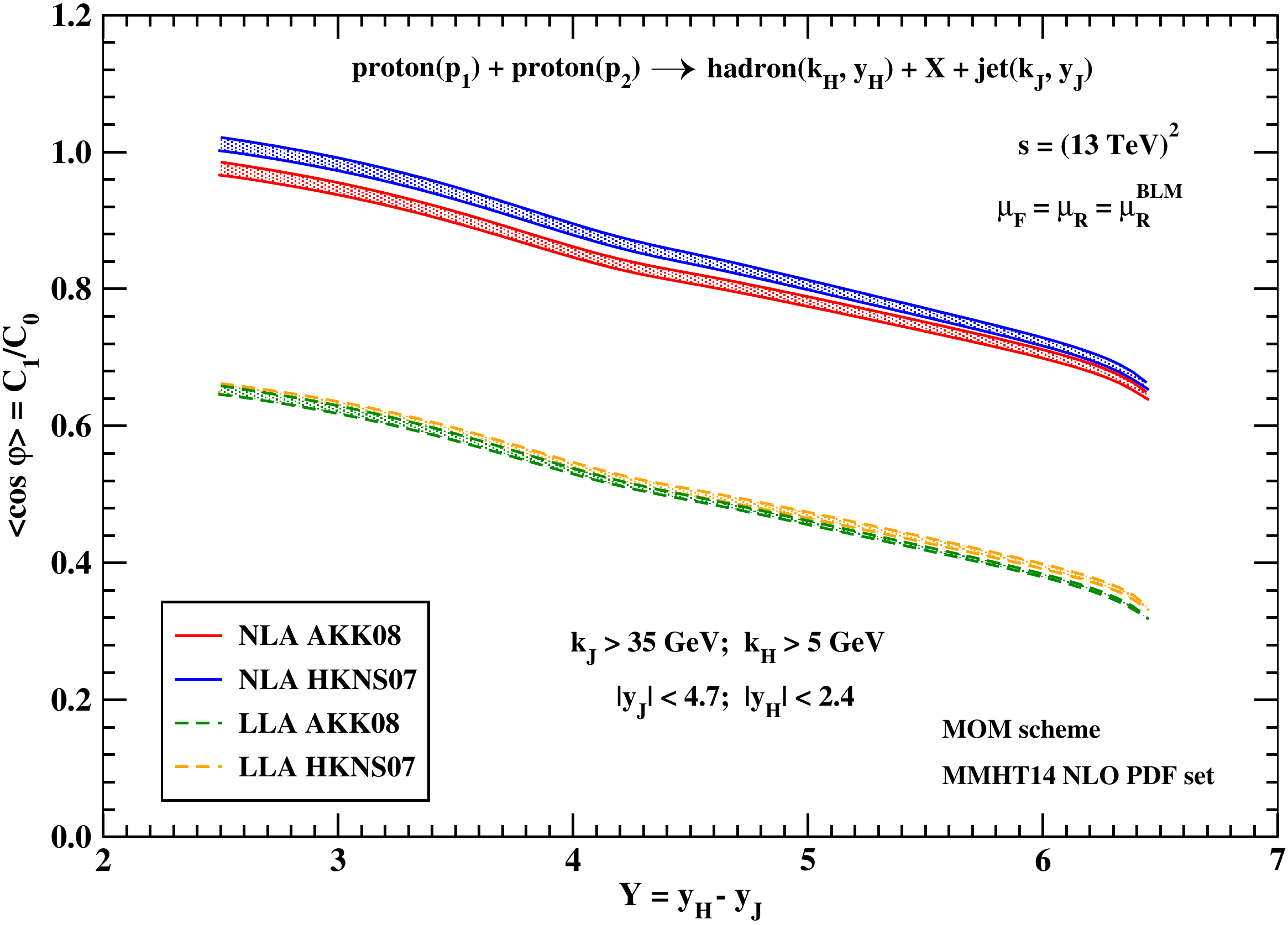}
  \hspace{0.25cm}
  \includegraphics[scale=0.30,clip]{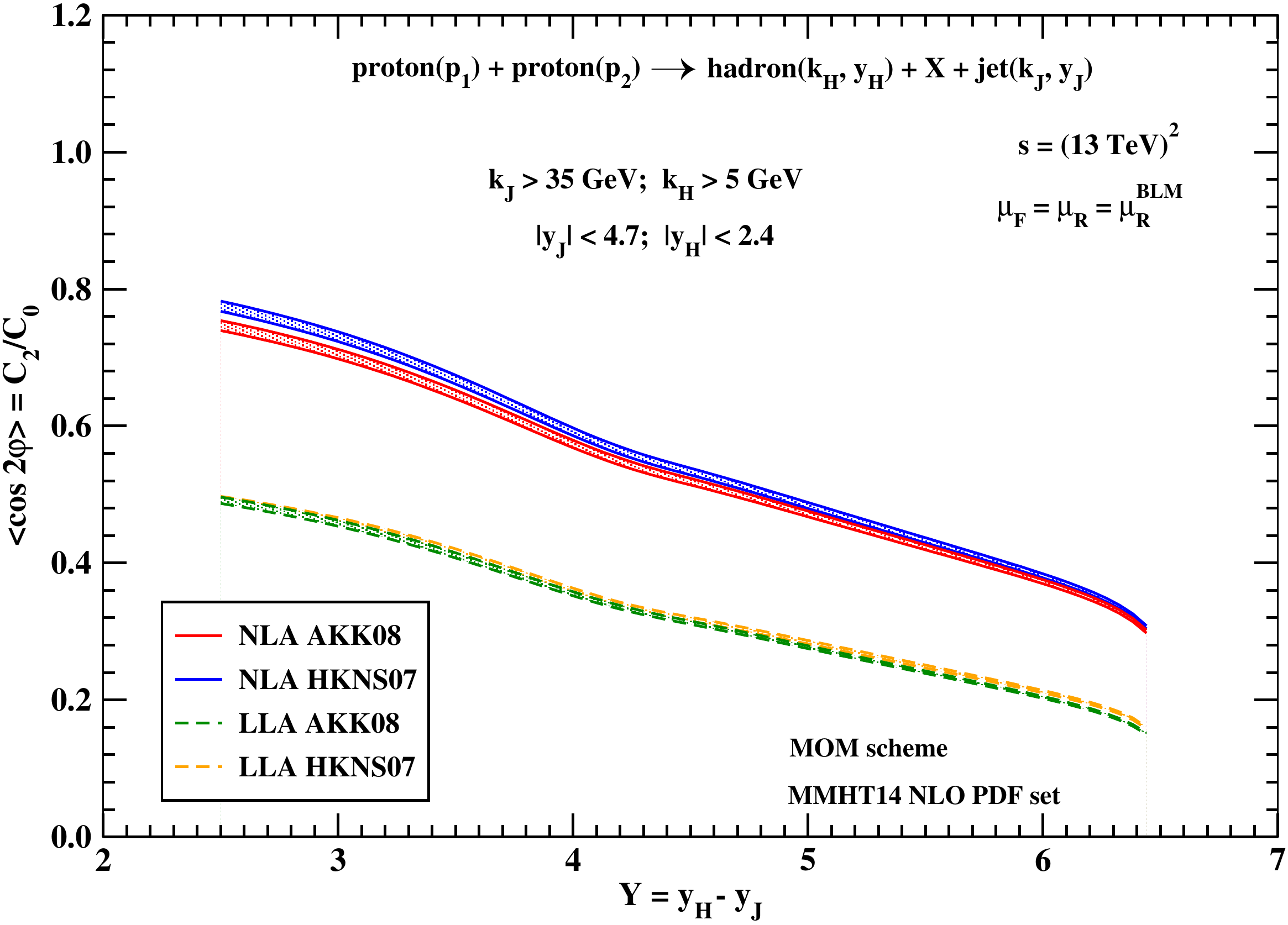}

  \includegraphics[scale=0.30,clip]{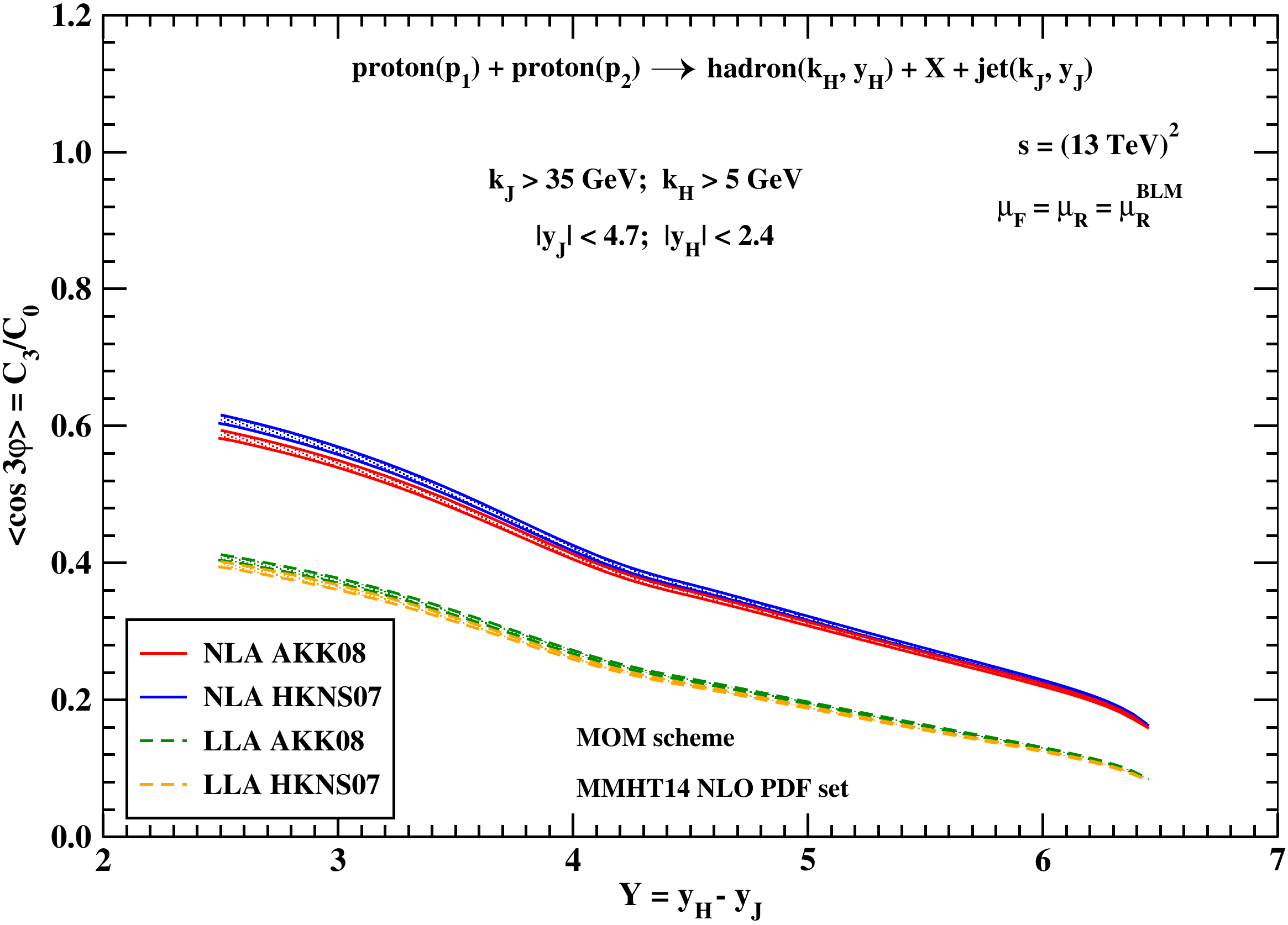}
  \hspace{0.25cm}
  \includegraphics[scale=0.30,clip]{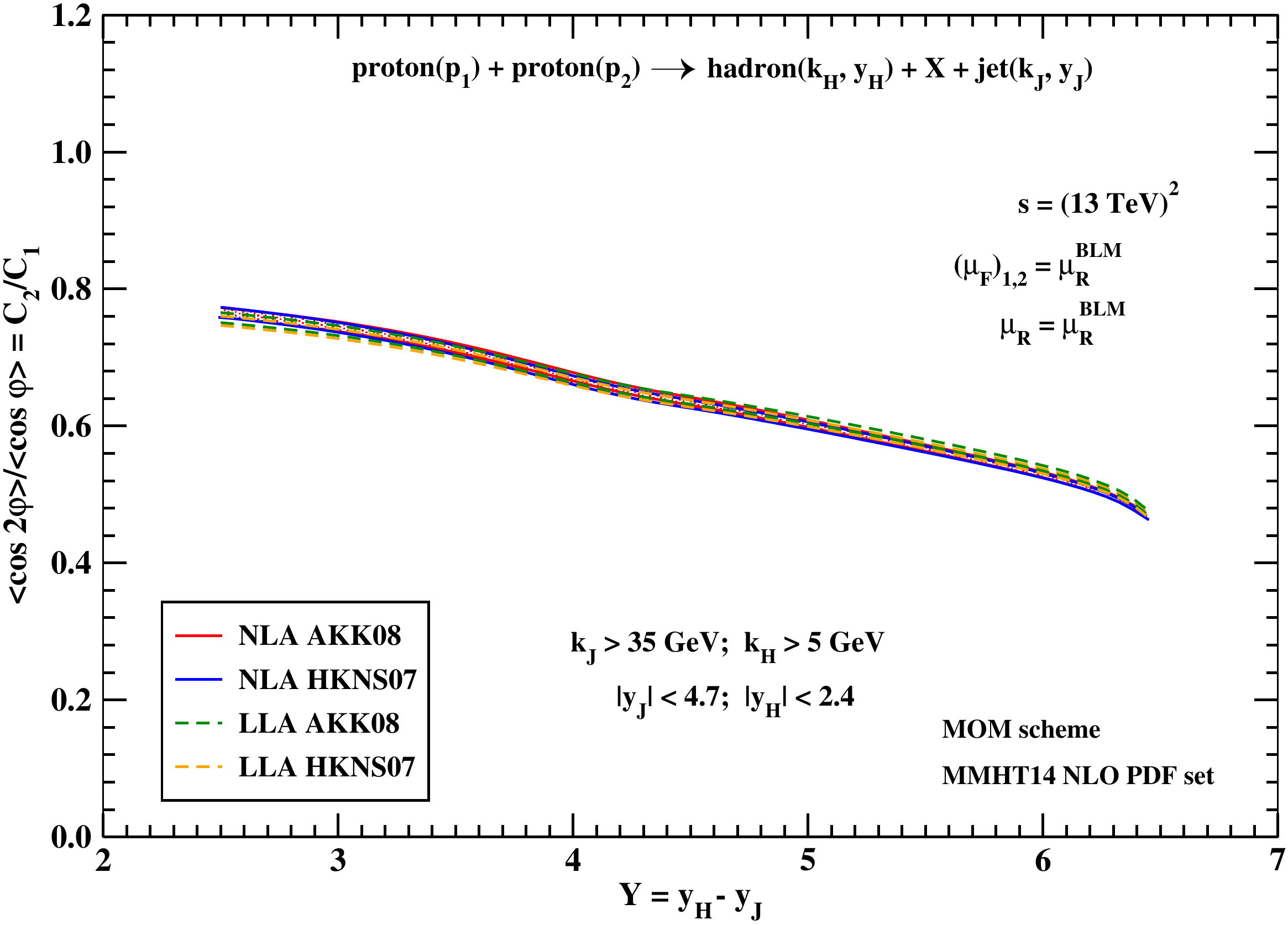}

  {\it a}) {\it CMS-jet} range.
  \vspace{0.5cm}
  
  \includegraphics[scale=0.31,clip]{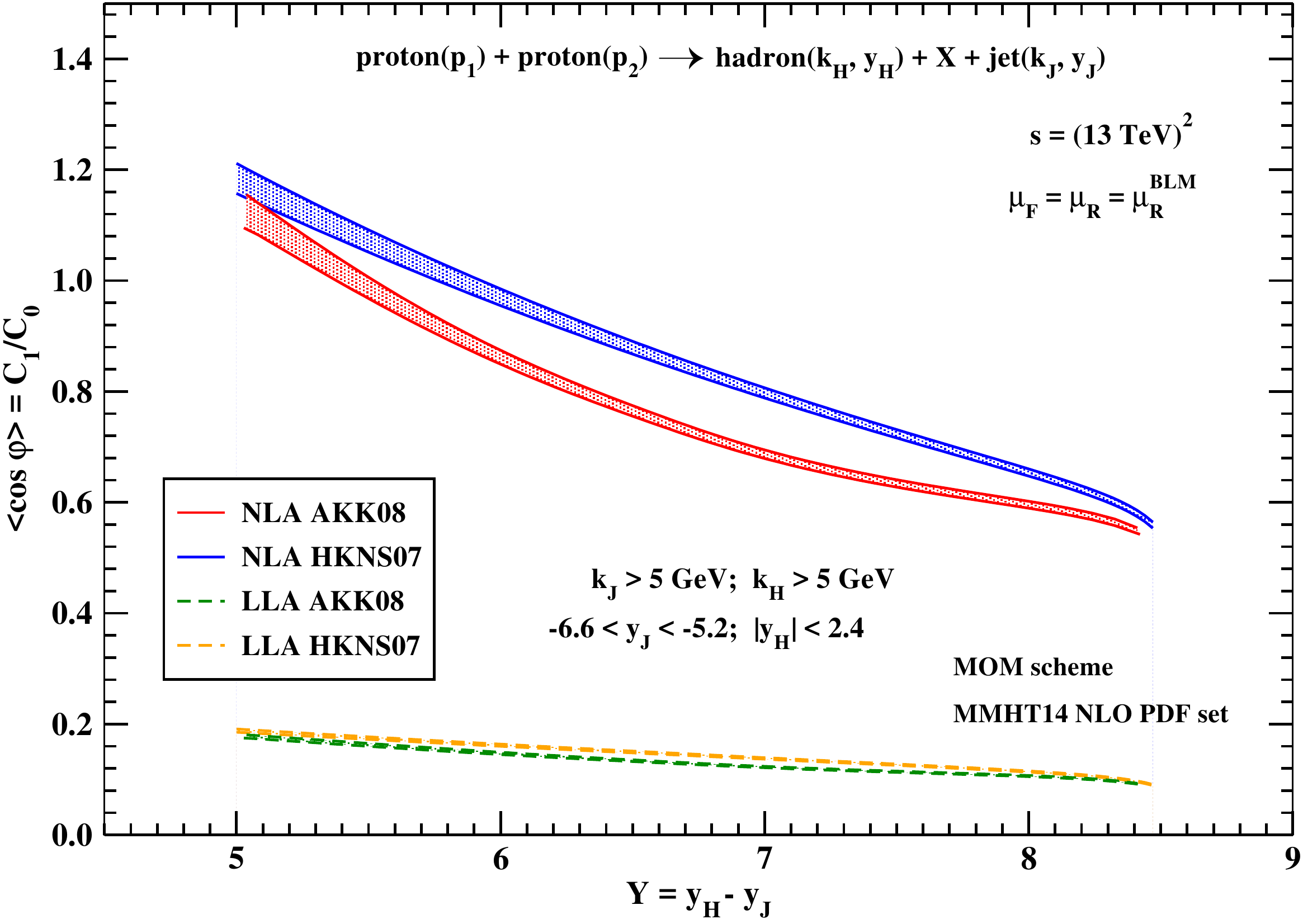}
  \hspace{0.25cm}
  \includegraphics[scale=0.31,clip]{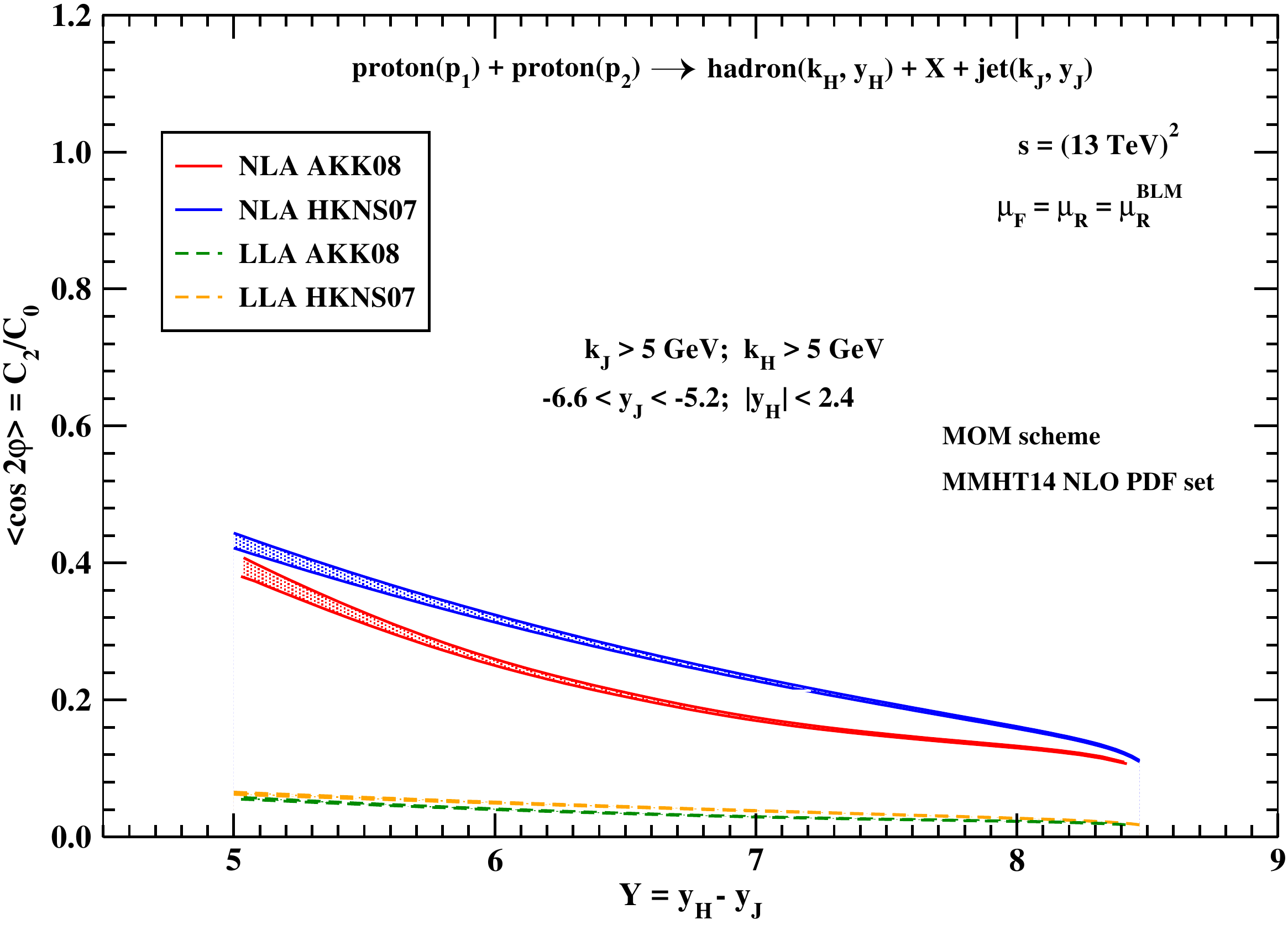}

  \includegraphics[scale=0.31,clip]{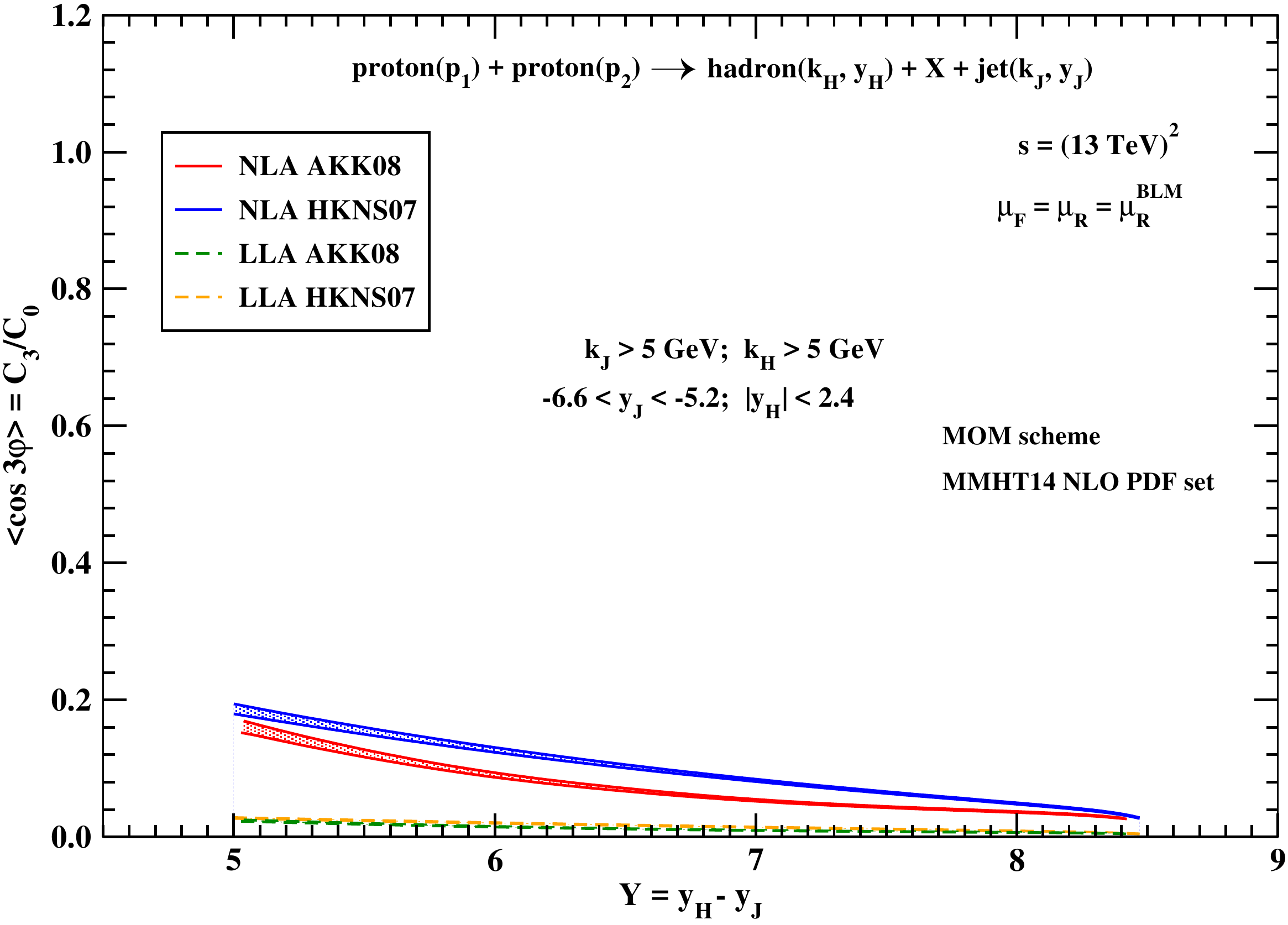}
  \hspace{0.25cm}
  \includegraphics[scale=0.31,clip]{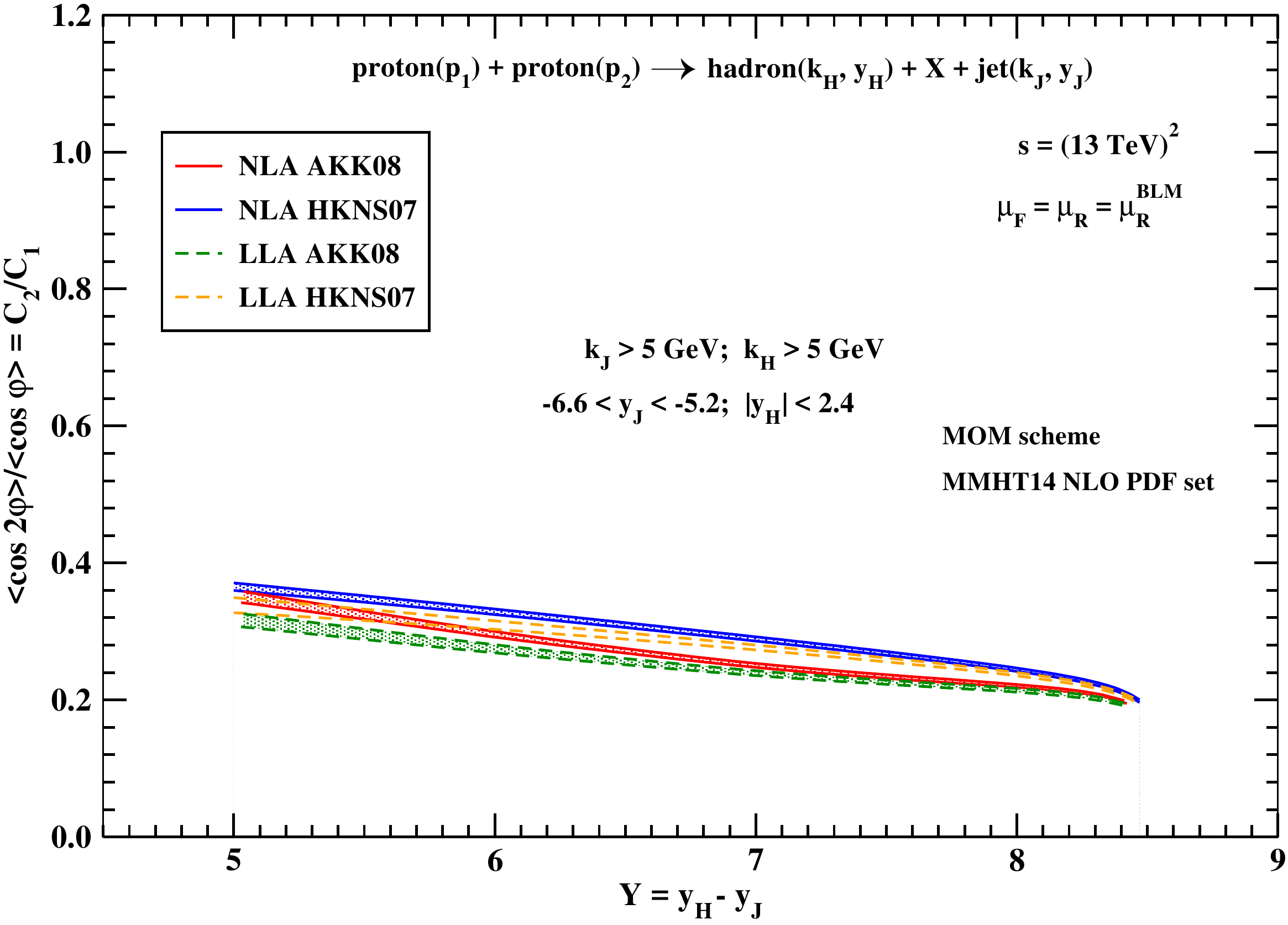}

  {\it b}) {\it CASTOR-jet} range.
  \vspace{0.25cm}

  \caption{$Y$-dependence of $R_{10}$, $R_{20}$, $R_{30}$ and $R_{21}$ in the two considered configurations, for $\sqrt{s} = 13$ TeV.}

  \label{fig:Cn_MOM_BLM_13}
\end{figure}

Two distinct final-state configuration are selected:
\begin{enumerate}
 \item
  \textit{\textbf{CMS-jet}} detection~\cite{Khachatryan:2016udy}: 
  both the hadron and the jet emitted inside the typical acceptances of the CMS detector:
  5 GeV $ < \kappa_H < $ 21.5 GeV, 35 GeV $ < \kappa_J < $ 60 GeV, 
  $|y_H| \leq 2.4$, 
  $|y_H| \leq 4.7$;
 \item
  \textit{\textbf{CASTOR-jet}}~\cite{CMS:2016ndp} detection: a hadron always tagged inside CMS, together with a very backward jet detected by CASTOR in the range: 5 GeV $ < \kappa_J \lesssim $ 17.68 GeV, 
  $-6.6 < y_J < -5.2$.
\end{enumerate}
In our calculations, done in the MOM renormalization scheme, the {\tt MMHT}~2014 NLO PDF parameterization~\cite{Harland-Lang:2014zoa}, together with two different NLO hadron FF sets, {\tt AKK}~2008~\cite{Albino:2008fy} and {\tt HKNS}~2007~\cite{Hirai:2007cx}, was employed. We used the Brodsky-Lepage-Mackenzie (BLM) scheme~\cite{Brodsky:1996sgBrodsky:1997sdBrodsky:1998knBrodsky:2002ka}, as given in its \emph{exact} version~\cite{Caporale:2015uva}, for the choice of the renormalization scale, $\mu_R$, and we set $\mu_F = \mu_R \equiv \mu_R^{\rm (BLM)}$.
In Fig.~\ref{fig:C0_comp_NLA_BLM_CMS} we examine $C_0$ for different inclusive NLA BFKL reactions: dijet (Mueller--Navelet), hadron-jet and dihadron production at $\sqrt{s} = 7$, 13 TeV in the {\it CMS-jet} range selection, whereas in Fig.~\ref{fig:Cn_MOM_BLM_13} we present results for different ratios $R_{nm} \equiv C_n/C_m$ at $\sqrt{s} =$ 13 TeV). For a comprehensive discussion on results, numerical tools and uncertainty estimate, we refer to Section 3 of~\cite{Bolognino:2018oth}.

\section{Conclusions and Outlook}
\vsb

We proposed a novel reaction as probe of the BFKL dynamics, {\it i.e.} the inclusive hadron-jet production at the LHC, giving predictions for cross section and azimuthal correlations in the NLA accuracy. The $Y$-dependence of our observables exhibits trends similar the ones found for semi-hard processes previously studied (\emph{e.g.} Mueller--Navelet process, inclusive dihadron production, etc...), when the jet is tagged inside the CMS detector, while new, unexpected features have appeared in the \emph{CASTOR-jet} configuration. A more detailed analysis in this direction, together with comparisons between BFKL-inspired and DGLAP-based, fixed-order calculations is underway.



\begin{thebibliography}{99}
\vsb

\bibitem{BFKL}
V.S.~Fadin, E.~Kuraev, L.~Lipatov, Phys. Lett. B \textbf{60}, 50 (1975);
%
Sov. Phys. JETP \textbf{44}, 443 (1976);
%
E.~Kuraev, L.~Lipatov, V.S.~Fadin, Sov. Phys. JETP \textbf{45}, 199 (1977);
%
I.~Balitsky, L.~Lipatov, Sov. J. Nucl. Phys. \textbf{28}, 822 (1978).

\bibitem{Gribov:1984tu}
  L.V.~Gribov, E.M.~Levin, M.G.~Ryskin,
  Phys.\ Rept.\  {\bf 100} (1983) 1.

\bibitem{Celiberto:2017ius}
  F.G.~Celiberto, PhD thesis,
  arXiv:1707.04315 [hep-ph].

\bibitem{Bolognino:2018rhb}
  A.D.~Bolognino, F.G.~Celiberto, D.Yu.~Ivanov, A.~Papa,
  Eur.\ Phys.\ J.\ C {\bf 78} (2018) no.12,  1023
  [arXiv:1808.02395 [hep-ph]].

\bibitem{Bolognino:2018mlw}
  A.D.~Bolognino, F.G.~Celiberto, D.Yu.~Ivanov, A.~Papa,
  arXiv:1808.02958 [hep-ph].

\bibitem{Bolognino:2019bko}
  A.D.~Bolognino, F.G.~Celiberto, D.Yu.~Ivanov, A.~Papa,
  arXiv:1902.04520 [hep-ph].


\bibitem{Ivanov:2004pp}
  D.Yu.~Ivanov, M.I.~Kotsky, A.~Papa,
  Eur.\ Phys.\ J.\ C {\bf 38} (2004) 195
  [hep-ph/0405297].

\bibitem{Ivanov:2005gn}
  D.Yu.~Ivanov, A.~Papa,
  Nucl.\ Phys.\ B {\bf 732} (2006) 183
  [hep-ph/0508162].

\bibitem{Ivanov:2006gt}
  D.Yu.~Ivanov, A.~Papa,
  Eur.\ Phys.\ J.\ C {\bf 49} (2007) 947
  [hep-ph/0610042].

\bibitem{Enberg:2005eq}
  R.~Enberg, B.~Pire, L.~Szymanowski, S.~Wallon,
  Eur.\ Phys.\ J.\ C {\bf 45} (2006) 759
   Erratum: [Eur.\ Phys.\ J.\ C {\bf 51} (2007) 1015]
  [hep-ph/0508134].

\bibitem{Mueller:1986ey}
A.H.~Mueller, H.~Navelet, 
Nucl. Phys. B \textbf{282}, 727 (1987).

\bibitem{Colferai:2010wu}
  D.~Colferai, F.~Schwennsen, L.~Szymanowski, S.~Wallon,
  JHEP {\bf 1012} (2010) 026
  [arXiv:1002.1365 [hep-ph]].

\bibitem{Caporale:2012ih}
  F.~Caporale, D.Yu.~Ivanov, B.~Murdaca, A.~Papa,
  Nucl.\ Phys.\ B {\bf 877} (2013) 73
  [arXiv:1211.7225 [hep-ph]].

\bibitem{Ducloue:2013wmi}
  B.~Duclou\'e, L.~Szymanowski, S.~Wallon,
  JHEP {\bf 1305} (2013) 096
  [arXiv:1302.7012 [hep-ph]].

\bibitem{Ducloue:2013bva}
  B.~Duclou\'e, L.~Szymanowski, S.~Wallon,
  Phys.\ Rev.\ Lett.\  {\bf 112} (2014) 082003
  [arXiv:1309.3229 [hep-ph]].
  
\bibitem{Caporale:2013uva}
  F.~Caporale, B.~Murdaca, A.~Sabio Vera, C.~Salas,
  Nucl.\ Phys.\ B {\bf 875} (2013) 134
  [arXiv:1305.4620 [hep-ph]].

\bibitem{Ducloue:2014koa}
  B.~Duclou\'e, L.~Szymanowski, S.~Wallon,
  Phys.\ Lett.\ B {\bf 738} (2014) 311
  [arXiv:1407.6593 [hep-ph]].

\bibitem{Caporale:2014gpa}
  F.~Caporale, D.Yu.~Ivanov, B.~Murdaca, A.~Papa,
  Eur.\ Phys.\ J.\ C {\bf 74}, no. 10, 3084 (2014)
  [Eur.\ Phys.\ J.\ C {\bf 75}, no. 11, 535 (2015)]
  [arXiv:1407.8431 [hep-ph]].

\bibitem{Ducloue:2015jba}
  B.~Duclou\'e, L.~Szymanowski and S.~Wallon,
  Phys.\ Rev.\ D {\bf 92} (2015) no.7,  076002
  [arXiv:1507.04735 [hep-ph]].

\bibitem{Caporale:2015uva}
  F.~Caporale, D.Yu.~Ivanov, B.~Murdaca, A.~Papa,
  Phys.\ Rev.\ D {\bf 91} (2015) no.11,  114009
  [arXiv:1504.06471 [hep-ph]].
  
\bibitem{Celiberto:2015yba}
  F.G.~Celiberto, D.Yu.~Ivanov, B.~Murdaca, A.~Papa,
  Eur.\ Phys.\ J.\ C {\bf 75} (2015) no.6,  292
  [arXiv:1504.08233 [hep-ph]].

\bibitem{Celiberto:2015mpa}
  F.G.~Celiberto, D.Yu.~Ivanov, B.~Murdaca and A.~Papa,
  Acta Phys.\ Polon.\ Supp.\  {\bf 8} (2015) 935
  [arXiv:1510.01626 [hep-ph]].

\bibitem{Celiberto:2016ygs}
  F.G.~Celiberto, D.Yu.~Ivanov, B.~Murdaca, A.~Papa,
  Eur.\ Phys.\ J.\ C {\bf 76} (2016) no.4,  224
  [arXiv:1601.07847 [hep-ph]].

\bibitem{Caporale:2018qnm}
  F.~Caporale, F.G.~Celiberto, G.~Chachamis, D.~Gordo G{\'o}mez, A.~Sabio Vera,
  Nucl.\ Phys.\ B {\bf 935} (2018) 412
  [arXiv:1806.06309 [hep-ph]].

\bibitem{Chachamis:2015crx}
  G.~Chachamis,
  arXiv:1512.04430 [hep-ph].

\bibitem{Ivanov:2012iv}
  D.Yu.~Ivanov, A.~Papa,
  JHEP {\bf 1207} (2012) 045
  [arXiv:1205.6068 [hep-ph]].
  
\bibitem{Celiberto:2016hae}
  F.G.~Celiberto, D.Yu.~Ivanov, B.~Murdaca, A.~Papa,
  Phys.\ Rev.\ D {\bf 94} (2016) no.3,  034013
  [arXiv:1604.08013 [hep-ph]].

\bibitem{Celiberto:2017ptm}
  F.~G.~Celiberto, D.Yu.~Ivanov, B.~Murdaca and A.~Papa,
  Eur.\ Phys.\ J.\ C {\bf 77} (2017) no.6,  382
  [arXiv:1701.05077 [hep-ph]].

\bibitem{Caporale:2015vya}
  F.~Caporale, G.~Chachamis, B.~Murdaca, A.~Sabio Vera,
  Phys.\ Rev.\ Lett.\  {\bf 116} (2016) no.1,  012001
  [arXiv:1508.07711 [hep-ph]].

\bibitem{Caporale:2015int}
  F.~Caporale, F.G.~Celiberto, G.~Chachamis, A.~Sabio Vera,
  Eur.\ Phys.\ J.\ C {\bf 76} (2016) no.3,  165
  [arXiv:1512.03364 [hep-ph]].

\bibitem{Caporale:2016soq}
  F.~Caporale, F.G.~Celiberto, G.~Chachamis, D.~Gordo.~G{\'o}mez, A.~Sabio Vera,
  Nucl.\ Phys.\ B {\bf 910} (2016) 374
  [arXiv:1603.07785 [hep-ph]].

\bibitem{Caporale:2016xku}
  F.~Caporale, F.G.~Celiberto, G.~Chachamis, D.~Gordo~G{\'o}mez, A.~Sabio Vera,
  Eur.\ Phys.\ J.\ C {\bf 77} (2017) no.1,  5
  arXiv:1606.00574 [hep-ph].

\bibitem{Celiberto:2016vhn}
  F.G.~Celiberto,
  Frascati Phys.\ Ser.\  {\bf 63} (2016) 43
  [arXiv:1606.07327 [hep-ph]].

\bibitem{Caporale:2016pqe}
  F.~Caporale, F.G.~Celiberto, G.~Chachamis, D.~Gordo G{\'o}mez, B.~Murdaca, A.~Sabio Vera,
  JCEGI 5 (2017) no.2, 47
  [arXiv:1610.04765 [hep-ph]].

\bibitem{Caporale:2016zkc}
  F.~Caporale, F.G.~Celiberto, G.~Chachamis, D.~Gordo~G{\'o}mez, A.~Sabio Vera,
  Phys.\ Rev.\ D {\bf 95} (2017) no.7,  074007
  [arXiv:1612.05428 [hep-ph]].

\bibitem{Celiberto:2017nyx}
  F.G.~Celiberto, D.Yu.~Ivanov, B.~Murdaca, A.~Papa,
  Phys.\ Lett.\ B {\bf 777} (2018) 141
  [arXiv:1709.10032 [hep-ph]];
%
  A.D.~Bolognino, F.G.~Celiberto, M.~Fucilla, D.Yu.~Ivanov, B.~Murdaca, A.~Papa,
  PoS(DIS2019)067
  [arXiv:1906.05940 [hep-ph]].

\bibitem{Boussarie:2017oae}
  R.~Boussarie, B.~Duclou\'e, L.~Szymanowski, S.~Wallon,
  Phys.\ Rev.\ D {\bf 97} (2018) no.1,  014008
  [arXiv:1709.01380 [hep-ph]].

\bibitem{Motyka:2014lya}
  L.~Motyka, M.~Sadzikowski, T.~Stebel,
  JHEP {\bf 1505} (2015) 087
  [arXiv:1412.4675 [hep-ph]].

\bibitem{Brzeminski:2016lwh}
  D.~Brzeminski, L.~Motyka, M.~Sadzikowski, T.~Stebel,
  JHEP {\bf 1701} (2017) 005
  [arXiv:1611.04449 [hep-ph]].

\bibitem{Celiberto:2018muu}
  F.G.~Celiberto, D.~Gordo G{\'o}mez, A.~Sabio Vera,
  Phys.\ Lett.\ B {\bf 786} (2018) 201
  [arXiv:1808.09511 [hep-ph]].

\bibitem{Golec-Biernat:2018kem}
  K.~Golec-Biernat, L.~Motyka, T.~Stebel,
  JHEP {\bf 1812} (2018) 091
  [arXiv:1811.04361 [hep-ph]].

\bibitem{Deak:2018obv}
  M.~Deak, A.~van Hameren, H.~Jung, A.~Kusina, K.~Kutak, M.~Serino,
  Phys.\ Rev.\ D {\bf 99} (2019) no.9,  094011
  [arXiv:1809.03854 [hep-ph]].

\bibitem{Bolognino:2018oth}
  A.D.~Bolognino, F.G.~Celiberto, D.Yu.~Ivanov, M.M.A.~Mohammed, A.~Papa,
  Eur.\ Phys.\ J.\ C {\bf 78} (2018) no.9,  772
  [arXiv:1808.05483 [hep-ph]].

\bibitem{Bolognino:2019yqj}
  A.D.~Bolognino, F.G.~Celiberto, D.Yu.~Ivanov, M.M.A.~Mohammed, A.~Papa,
  arXiv:1902.04511 [hep-ph].

\bibitem{Khachatryan:2016udy}
  V.~Khachatryan {\it et al.} [CMS Collaboration],
  JHEP {\bf 1608} (2016) 139   [arXiv:1601.06713 [hep-ex]].

\bibitem{CMS:2016ndp}
  CMS Collaboration [CMS Collaboration],
  CMS-PAS-FSQ-16-003.

\bibitem{Harland-Lang:2014zoa}
  L.A.~Harland-Lang, A.D.~Martin, P.~Motylinski, R.S.~Thorne,
  Eur.\ Phys.\ J.\ C {\bf 75} (2015) no.5,  204
  [arXiv:1412.3989 [hep-ph]].

\bibitem{Albino:2008fy}
  S.~Albino, B.A.~Kniehl, G.~Kramer,
  Nucl.\ Phys.\ B {\bf 803}, 42 (2008).
  [arXiv:0803.2768 [hep-ph]].

\bibitem{Hirai:2007cx}
  M.~Hirai, S.~Kumano, T.-H.~Nagai, K.~Sudoh,
  Phys.\ Rev.\ D {\bf 75}, 094009 (2007). 
  [hep-ph/0702250].

\bibitem{Brodsky:1996sgBrodsky:1997sdBrodsky:1998knBrodsky:2002ka}
  S.J.~Brodsky, F.~Hautmann, D.E.~Soper, Phys. Rev. Lett. {\bf 78}, 803
  (1997). [Erratum: Phys. Rev. Lett. {\bf 79}, 3544 (1997)];
  %
  Phys. Rev. D {\bf 56}, 6957 (1997);
  %
  S.J.~Brodsky, V.S.~Fadin, V.T.~Kim, L.N.~Lipatov, G.B.~Pivovarov, 
  JETP Lett. {\bf 70}, 155 (1999);
  %
  JETP Lett. {\bf 76}, 249 (2002).

\end{thebibliography}
\end{document}